\documentclass{article}

\usepackage[final, nonatbib]{neurips_2023}

\usepackage[utf8]{inputenc} 
\usepackage[T1]{fontenc}    
\usepackage{hyperref}       
\usepackage{url}            
\usepackage{booktabs}       
\usepackage{amsfonts}       
\usepackage{nicefrac}       
\usepackage{microtype}      
\usepackage{xcolor}         
\usepackage{amssymb}
\usepackage{amsmath}
\usepackage{mathtools}
\usepackage{graphicx}
\usepackage[backend=biber, style=numeric, maxnames=99]{biblatex}
\usepackage[backend=biber,style=numeric]{biblatex}
\title{Approximation of Intractable Likelihood Functions in Systems Biology via Normalizing Flows}

\author{%
  Vincent D. Zaballa  \\
  Department of Biomedical Engineering\\
  University of California, Irvine\\
  Irvine, CA 92697 \\
  \texttt{vzaballa@uci.edu} \\
  \And
  Elliot E. Hui \\
  Department of Biomedical Engineering \\
  University of California, Irvine\\
  Irvine, CA 92697 \\
  \texttt{eehui@uci.edu} \\
}

\begin{document}

\maketitle

\begin{abstract}
  Systems biology relies on mathematical models that often involve complex and intractable likelihood functions, posing challenges for efficient inference and model selection. Generative models, such as normalizing flows, have shown remarkable ability in approximating complex distributions in various domains. However, their application in systems biology for approximating intractable likelihood functions remains unexplored. Here, we elucidate a framework for leveraging normalizing flows to approximate complex likelihood functions inherent to systems biology models. By using normalizing flows in the Simulation-based inference setting, we demonstrate a method that not only approximates a likelihood function but also allows for model inference in the model selection setting. We showcase the effectiveness of this approach on real-world systems biology problems, providing practical guidance for implementation and highlighting its advantages over traditional computational methods. 
\end{abstract}


\section{Introduction}
\label{sec:introduction}

Systems biology aims to understand the complex interactions within biological systems by modeling them as integrated networks of genes, proteins, and biochemical reactions. Analogous to ensemble methods in machine learning, where individual models are integrated to improve predictive accuracy, systems biology aims to model complex biological systems as integrated wholes to provide deeper insights into their function and behavior. Leveraging computational approaches, including machine learning algorithms for network analysis and high-dimensional data interpretation, systems biology offers a paradigm shift from reductionist methods that examine individual components, such as proteins, in isolation. 

A pressing challenge in systems biology is the problem of inference in models characterized by intractable likelihood functions. These complex models often encapsulate the high dimensionality and non-linearity innate to biological systems but come at the cost of computational intractability for standard statistical methods. Conventional approaches, such as maximum likelihood estimation via least squares regression, are generally ill-suited for these models, failing to provide robust parameter estimates that can be reliably used for further analysis or uncertainty quantification. This creates a bottleneck that limits our ability to draw meaningful insights from biological networks. Examples of biological networks with intractable likelihoods are those governed by mass action kinetics, including protein dynamics and metabolic systems \cite{feinberg2019foundations}.

Simulation-based inference (SBI) methods \cite{Cranmer2020}, particularly those employing normalizing flows, have begun to bridge the gap in tackling the challenging problem of approximating intractable likelihoods in systems biology. Normalizing flows transform a complex or unknown distribution into a simple, tractable one, such as a Gaussian, effectively learning the data's underlying distribution without requiring an explicit likelihood function. This offers a powerful tool for probabilistic modeling of complex systems, allowing for both robust inference via posterior predictive checks and uncertainty quantification for each model parameter, as well as model comparison.

A salient challenge within systems biology is the inference of network parameters pertaining to the Bone Morphogenetic Protein (BMP) signaling pathway \cite{antebi2017combinatorial}. This pathway assumes a pivotal role in both pathological and developmental contexts, being implicated in an array of malignancies, such as cancer, as well as critical developmental processes \cite{salazar2016bmp}. Given its dichotomous functions, understanding the intricacies of BMP and other signaling pathways becomes imperative for predicting their responses to external modulations, such as pharmacological perturbations.

Prior studies modeled the BMP pathway using mass action kinetics that qualitatively described observed experimental outputs \cite{antebi2017combinatorial}. However, traditional parameter fitting methods such as least square regression yield an expansive array of plausible parameter values without a probabilistic interpretation, thereby aggravating the issue of model uncertainty. Additionally, the BMP pathway can be captured by multiple competing models \cite{su2022ligand}, introducing the need for effective model discrimination. Traditional criteria for model selection, such as the Akaike Information Criterion (AIC) and the Bayesian Information Criterion (BIC), offer point estimates of model probabilities. While these methods are generally efficacious for models with tractable likelihoods, they encounter limitations when applied to models characterized by intractable likelihood functions \cite{mackay2003information}. Hence, there is a need for new computational techniques that can conduct reliable inference and identify the most probable fitting model in the context of observed data, circumventing the limitations imposed by intractable likelihoods.

We demonstrate the first application of SBI techniques that use normalizing flows to address the task of approximating likelihood functions for the BMP signaling pathway. Using experimental data \cite{klumpe2022context}, we address the challenges associated with inferring these likelihood functions and demonstrate how to infer likelihoods of high-dimensional observed data \textit{and} high dimensional simulator parameters using statistical features of i.i.d. likelihoods to infer a data space $\mathcal{D}$ and parameter space $\mathcal{P}$, such that \(\mathcal{D} \in \mathbb{R}^{940}, \mathcal{P} \in \mathbb{R}^{m}\) where \(m \leq 70\). Using these approximated likelihoods, we not only scrutinize the uncertainty enveloping each model's parameters but also offer indications regarding which model is more plausible given the observed data. Our findings underscore the efficacy of deep generative models, such as normalizing flows, in the context of biological model parameter inference and model selection. Finally, these surrogate likelihoods hold the potential to be integrated into Bayesian Optimal Experimental Design (BOED) frameworks \cite{Foster2019a, kleinegesse2019efficient}, thereby minimizing both the temporal and experimental resources required to achieve a model with robust predictive accuracy in biology.


\section{Background}
\label{sec:background}

\textbf{The Intractable Likelihood of the BMP Model.} The BMP signaling pathway is commonly described by mass action kinetics and conservation laws, capturing the dynamics of downstream genetic expression signals reaching a steady-state. Two principal models exist for describing this system: the ``twostep'' model which involves equations \(A_i + L_j \leftrightarrow D_{ij}\) and \(D_{ij} + B_k \leftrightarrow T_{ijk}\), and the ``onestep'' model represented by \(A_i + B_k + L_j \leftrightarrow T_{ijk}\), where the binding affinity $K$ represents the transition from one state to another. Additionally, $i,j,k$ represent the $i^\textit{th}$ type $A$ receptor, the $j^\textit{th}$ protein ligand complex, and $k^\textit{th}$ type $B$ receptor, $D$ represents a dimeric protein complex, and $T$ represents a trimeric complex. Both models then describe the phosphorylation efficiency, $\varepsilon$, of a trimeric complex $T$ sending a downstream gene expression signal, $S$, as $S = \varepsilon T$. While these models are originally formulated as ordinary differential equations (ODEs), they become differential algebraic equations under certain conditions such as fixed volume and large ligand concentration. These models can be solved via convex optimization, however, this results in an intractable likelihood functions. The lack of a likelihood function for the BMP pathway requires novel computational methods, such as SBI, for further refinement and understanding of models' parameters and probabilities given observed data, $x_o$.

\textbf{Normalizing Flows.} Normalizing flows constitute a specialized subset of invertible and differentiable neural networks engineered to represent a series of monotonic transformations. These transformations are designed to minimize the divergence between a base distribution, commonly Gaussian, denoted as \( p_u(\mathbf{u}) \), and the target data distribution, \( p_x(\mathbf{x}) \). This is formulated using the change of variables formula and a composition of monotonic diffeomorphic functions, \( \mathbf{f}_{\phi} \), which can be parameterized via neural networks. The formal transformation from the base distribution to the data distribution is:

\begin{equation}
    p_x(\mathbf{x})  = p_u(f^{-1}_{\phi} (\mathbf{x})) \left\lvert \det \frac{\partial \mathbf{f}^{-1}_{\phi}}{\partial \mathbf{x}} \right\rvert.
\end{equation}

A thorough review on the theoretical framework and practical application of normalizing flows can be found in \cite{Kobyzev_2021} and \cite{Papamakarios2019}. In this work, we use normalizing flows to model the intractable likelihood $p_\phi(x|\theta)$, where $\phi$ are parameters of the normalizing flow and $\theta$ represent biological model parameters of interest, such as binding affinity, $K$, and phosphorylation efficiency, $\varepsilon$, such that $K, \varepsilon \in \theta$ and $x$ is simulated data from the simulator that we want to get close to the observed data, $x_o$. Using Bayes Theorem we can then infer parameters' posterior distribution as $p(\theta|x_o) \propto p_\phi(x_o|\theta)p(\theta)$.

\textbf{Simulation-based Inference.} Concurrently with advancements in normalizing flow architectures, significant advancements have been made in SBI algorithms for sequential posterior approximation \cite{greenberg2019automatic, Papamakarios2016, lueckmann2018likelihood, miller_contrastive_2022, Durkan2019, hermans2020likelihoodfree}. These SBI methodologies estimate various elements, such as the posterior distribution, likelihood, and the ratio of posterior to prior probabilities, to gauge the posterior of a specific model given observed data. SBI techniques are valuable in disciplines with complicated simulations based on mathematical models like particle physics \cite{brehmer2021simulation}, where functions are readily simulated but challenging to evaluate analytically.

These sequential SBI methods take inspiration from importance sampling to refine a posterior, likelihood, or likelihood ratio given an observed sample of interest, $x_o$. There are a variety of sequential SBI techniques, each with their own benefits, and reviewed in \cite{lueckmann2021benchmarking}. We employ the Sequential Neural Likelihood (SNL) approach \cite{Papamakarios2018}, which utilizes a normalizing flow to approximate the likelihood represented by a simulator and conditioned over multiple rounds on observed data. We chose to use likelihoods as the likelihood ratio can be used as a form of model comparison, which is another important desiderata of this paper.

\textbf{Model selection.} The Bayes Factor (BF) is another term for the ratio of model probabilities, expressed as \(BF = \frac{p(\mathcal{M}_0)}{p(\mathcal{M}_1)}\). It serves as a criterion for selecting between models. A \(BF > 10\) indicates strong support for \(\mathcal{M}_0\), while a \(BF < \frac{1}{10}\) reveals strong support for \(\mathcal{M}_1\). The BF inherently favor simpler models due to the Bayesian version of Occam's razor. This, however, depends on unbiased computation of the model's marginal likelihood. For more details on assorted model selection methods, refer to \cite{pml1Book}.

To engage in model selection, we need an estimate of the marginal likelihood for each model to compute the BF. One approach to achieve this is by sampling from a likelihood using posterior samples $p(x|\mathcal{M}) = \int_\Theta p(x | \theta, \mathcal{M})p(\theta|\mathcal{M}) d\theta \approx \frac{1}{N}\sum_{i=1}^N p(x | \theta_i, \mathcal{M})$. When each model's prior probabilities are equal, the BF reduces to a comparison of two model likelihoods such that $\frac{p(\mathcal{M}_1)}{p(\mathcal{M}_0)} = \frac{p(x|\mathcal{M}_1)}{p(x|\mathcal{M}_0)}$. We use this formulation in this paper given that both models evaluated are equally likely apriori.



\section{Experimental Evaluation}
\label{sec:experiments}

We employed two separate normalizing flow architectures for training both the onestep and twostep BMP models, utilizing the SNL method within the larger SBI framework. Specifically, we used the neural spline flow model \cite{Durkan2019}, wherein each model was constructed with five flow layers and four bins in each layer to approximate invertible polynomial functions. Each layer was conditioned on the model parameters, \( \theta \), as well as on static experimental design parameters previously employed, denoted as \( \xi \), which represents the concentration of BMP protein ligand administered ($L$) and the cell type, each with its own respective pseudo-concentrations of Type $A$ and $B$ receptors. This facilitated the creation of an approximate likelihood \( p_\phi(x|\theta | \xi) \) for incorporation into the SNL process. The training regimen extended over 9 rounds of SBI, during which we collected posterior parameters, gauged the median distance between simulated and observed data, and logged probabilities. In each round, 1,000 data points were simulated based on the posterior distribution from the previous round, starting with a uniform prior. Mini-batches comprised 100 data points each were used to train a normalizing flow on both the onestep and twostep models. All priors and subsequently simulated data were combined to form an aggregate training dataset for the normalizing flows. Concurrently, 5\% of the data was reserved in each round as a validation set for early stopping, which took place if the validation error did not decrease for more than 20 epochs. After each round, the approximate likelihood was used to infer a new posterior using $p_i(\theta|x_o) \propto p_\phi(x_o|\theta)p(\theta)$, where $p_i(\theta|x_o)$ represents the posterior after round $i$ using the trained likelihood from the previous round.

Given that the observed data point, \( x_o \), was a 940-dimensional output, modeling of the joint likelihood is computationally impractical. To address this issue, we exploited the property that the joint distribution could be factorized into a product of individual likelihoods, \( p_\phi(x_1, \ldots, x_{940}) = \prod_{i=1}^{940} p_\phi(x_i) \), thus simplifying the computational requirements for both training and the architecture.

\begin{figure}[t]
  \centering
  \vspace{-10pt}
  \includegraphics[width=\textwidth]{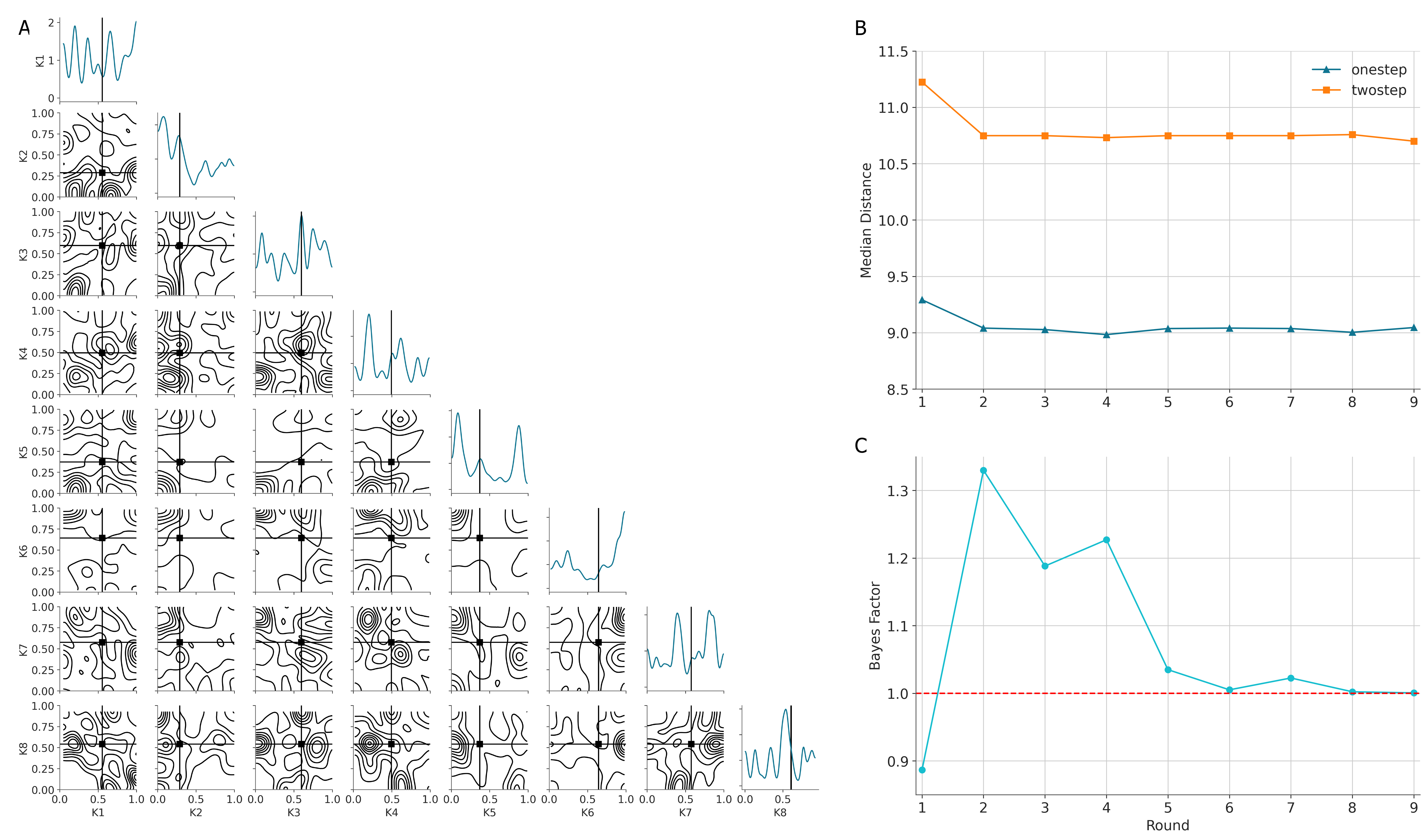}
  \caption{(A) Example of pairplots generated of the first 8 binding affinities, $K$, for the onestep model. Crosshairs indicate median values and occasionally align with a maximum a posteriori estimate of parameters' marginal distribution, but other times do not, reflecting uncertainty in the parameter values. (B) Median distance between $x_o$ and simulated values $x_i \sim p_i(x|\theta_{i-1})$ from the simulator using priors drawn from round $i-1$. (C) Bayes Factor of the onestep to twostep model using each model's respective density estimator in round $i$ and the final posterior returned after the last round. The dotted red line indicates the boundary between preference for the onestep model (>1) and the twostep model (<1). We see early indication that model 1 is slightly more likely, supported by the median distance, but diminishing probability with subsequent trained likelihoods, indicating ambiguity in model or overfitting of the likelihoods.
  \label{fig:pairplots}
  \vspace{-10pt}}
\end{figure}

Figure \ref{fig:pairplots} shows (left) a subset example of pairplots for the binding affinity parameters from the onestep model, (top right) median distance from $x_o$, and (bottom right) BF calculation using normalizing flows from each round. The final round's posterior is qualitatively different than the flat prior used in the initial round. The median distance serves as a validation metric for the normalizing flow's approximation of the true likelihood. The BF plot seems to agree with the median distance in earlier rounds that the onestep model seems to be a slightly better model for the data, but the convergence to essentially no difference in the later rounds indicates either that it is still ambiguous as to which model better fits the data, or, that the normalizing flows may not be expressive enough to accurately represent each likelihood.  Pairplots of all parameters for both models can be found in Appendix \ref{sec:appendix_a}.



\section{Discussion}
\label{sec:discussion}


We demonstrated how to use normalizing flows to estimate the likelihood of systems biology models, overcoming their known limitations with high-dimensional data \cite{10.7554/eLife.56261} by simplifying input dimensions and demonstrated utility in model selection. Future work will explore deeper flow networks and alternative methods like diffusion models, which are better at handling high-dimensional datasets while permitting a likelihood approximation via Probability Flow ODE \cite{song2021scorebased}.



\printbibliography[heading=bibnumbered]

\newpage
\appendix
\section{Full Posteriors of Onestep and Twostep Models}
\label{sec:appendix_a}

For qualitative reference, we present the entire posteriors for both binding affinities $K$, and phosphorylation efficiency $\varepsilon$, in this section. 

\begin{figure}[h]
  \centering
  \begin{minipage}[b]{0.45\textwidth}
    \centering
    \includegraphics[width=\linewidth]{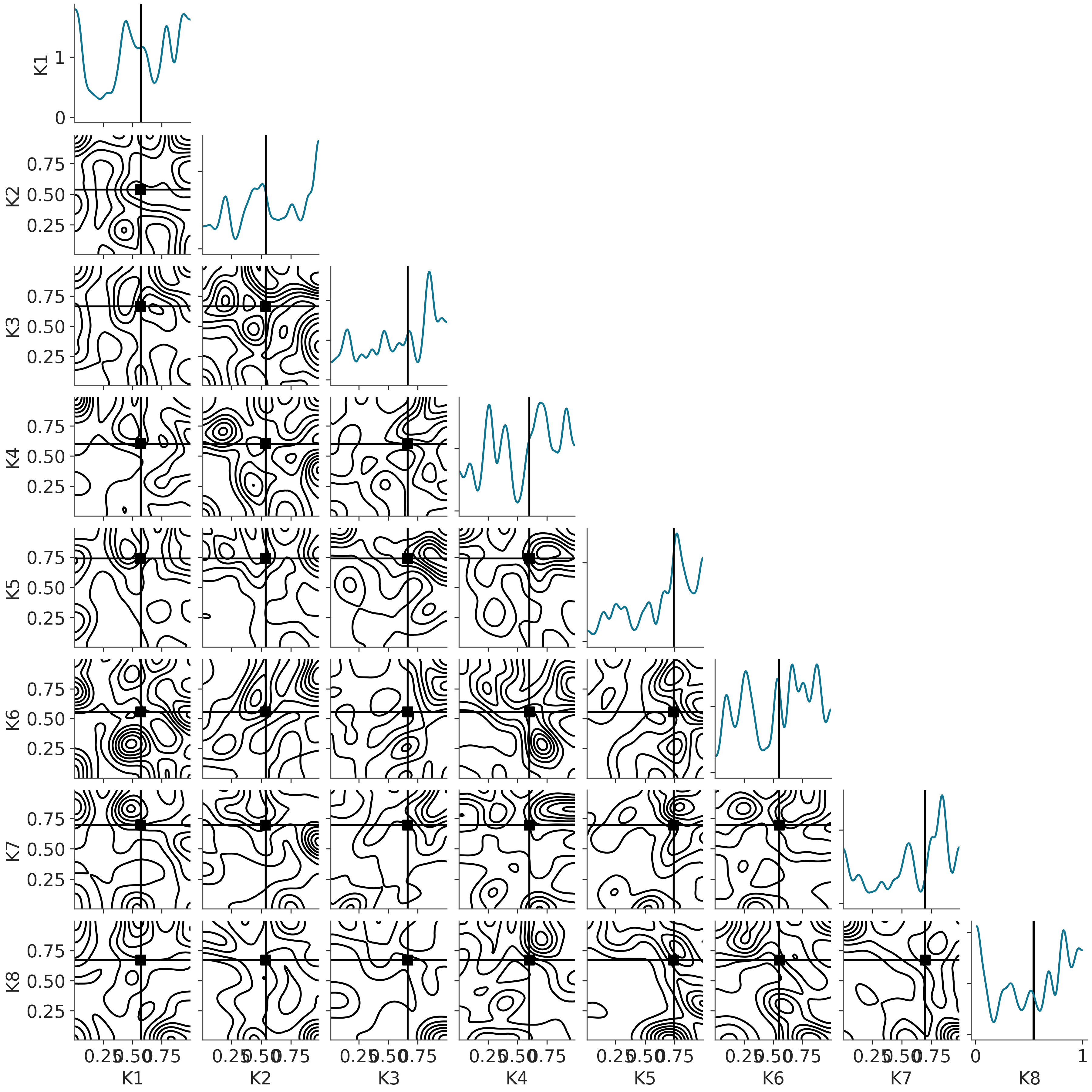}
    \label{fig:pairplots_K}
  \end{minipage}\hfill
  \begin{minipage}[b]{0.45\textwidth}
    \centering
    \includegraphics[width=\linewidth]{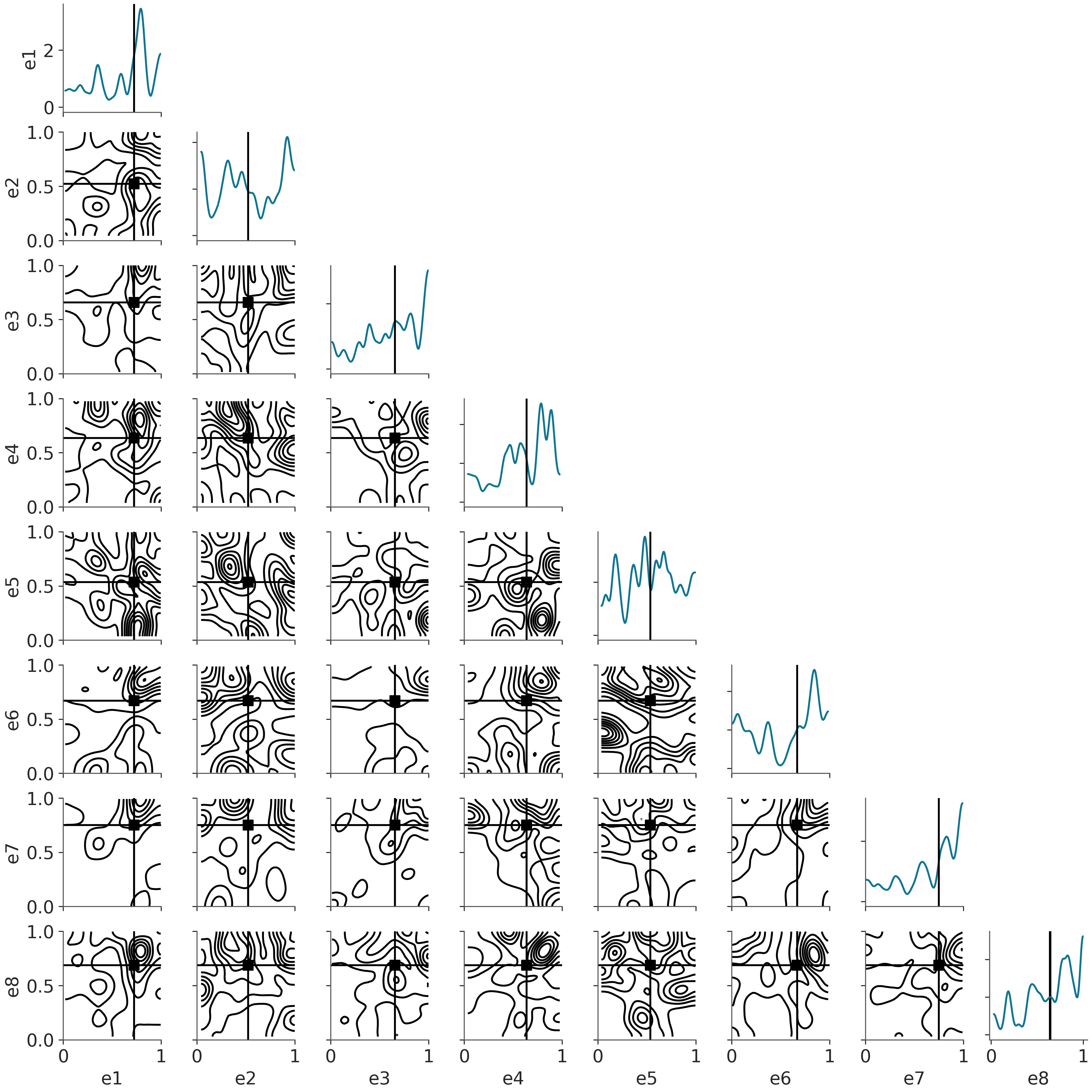}
    \label{fig:pairplots_e}
  \end{minipage}
  \caption{Pairplots for subsets of the \( K \) (left) and \( \varepsilon \) (right) parameters for the onestep model.}
  \label{fig:pairplots_onestep}
\end{figure}

\textbf{Onestep Model Posteriors.} We begin with a subset of posteriors for the onestep model shown in Figure \ref{fig:pairplots_onestep}. Generally, the posteriors for both types of parameters seem to have dominating peaks for both sets of parameters, which may indicate convergence of the model to those peak estimates. Occasionally, there are bimodal distributions for binding affinities, which physically means either the complex weakly or strongly prefers. Future experiments could help distinguish whether there is strong or weak binding affinity.

\begin{figure}[h]
  \centering
  \begin{minipage}[b]{0.45\textwidth}
    \centering
    \includegraphics[width=\linewidth]{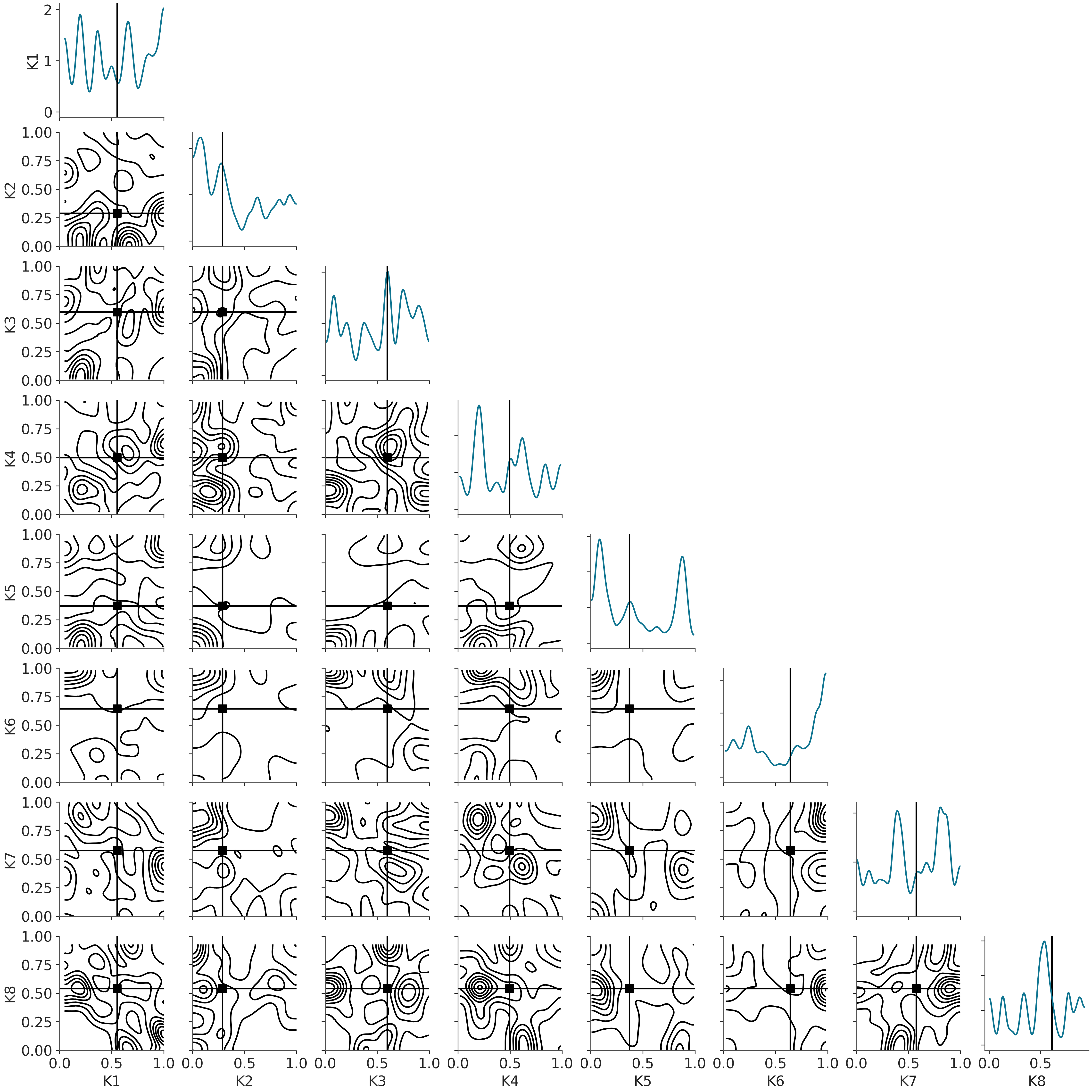}
    \label{fig:pairplots2_K}
  \end{minipage}\hfill
  \begin{minipage}[b]{0.45\textwidth}
    \centering
    \includegraphics[width=\linewidth]{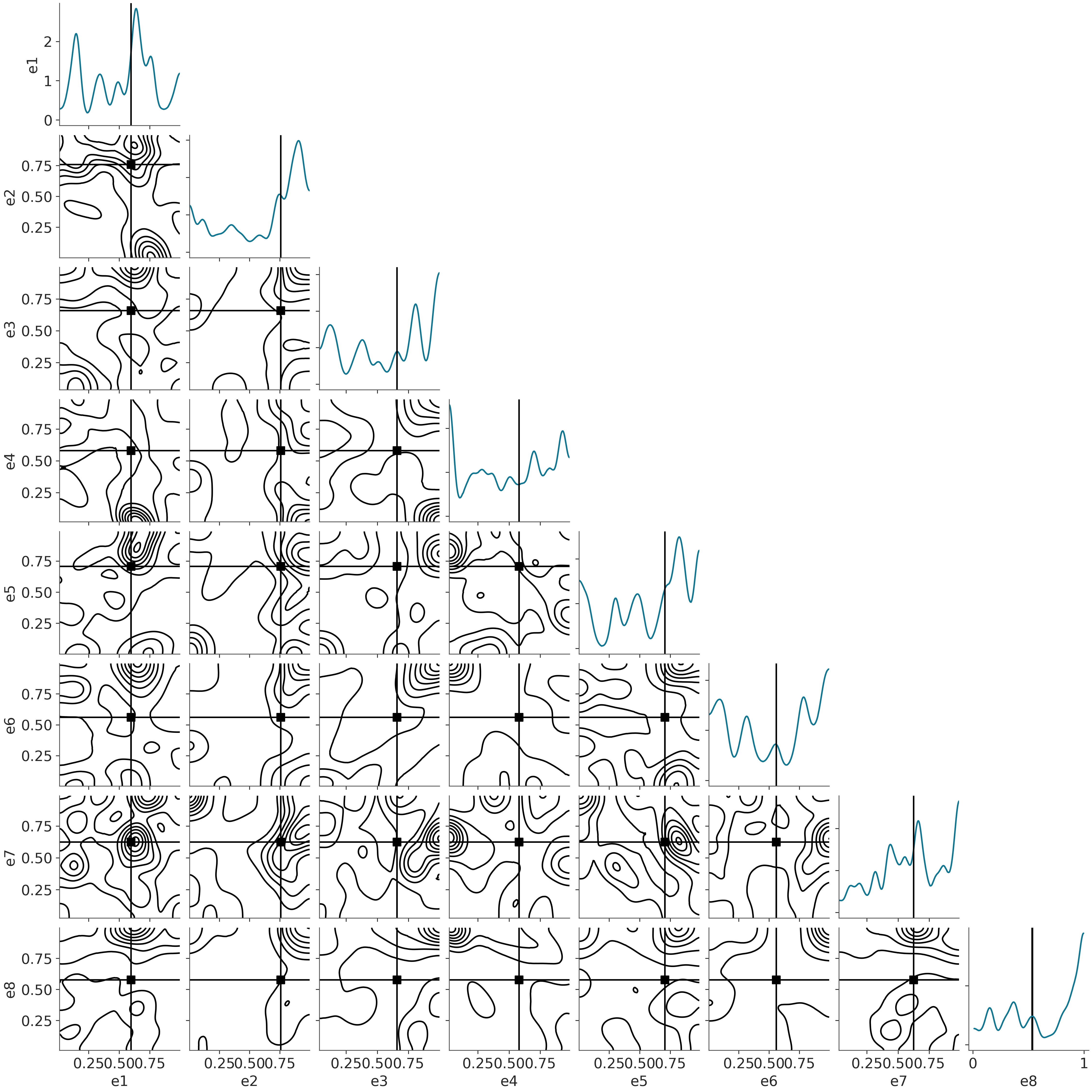}
    \label{fig:pairplots2_e}
  \end{minipage}
  \caption{Pairplots for subsets of the \( K \) (left) and \( \varepsilon \) (right) parameters for the twostep model.}
  \label{fig:pairplots_twostep}
\end{figure}

\textbf{Twostep Model Posteriors.} Looking at a subset of posteriors for the twostep model, Figure \ref{fig:pairplots_twostep} also shows uncertain posteriors but with more "peaks" in parameter values that could be possible. In a BOED context, this would indicate conducting experiments that could help elucidate between these peaks to drive convergence to one set of parameters.

\end{document}